# Providing Rate Guarantees to TCP over the ATM GFR Service[1]

Rohit Goyal, Raj Jain, Sonia Fahmy, and Bobby Vandalore
The Ohio State University Department of Computer and Information Science
Columbus, OH 43210-1277
Contact author E-mail: goyal@cis.ohio-state.edu
Phone: 614-688-4482, Fax: 614-292-2911

**Abstract:**

The ATM Guaranteed Frame Rate (GFR) service is intended for best effort traffic that can benefit from minimum throughput guarantees. Edge devices connecting LANs to an ATM network can use GFR to transport multiple TCP/IP connections over a single GFR VC. These devices would typically multiplex VCs into a single FIFO queue. It has been shown that in general, FIFO queuing is not sufficient to provide rate guarantees, and per-VC queuing with scheduling is needed. We show that under conditions of low buffer allocation, it is possible to control TCP rates with FIFO queuing and buffer management. We present analysis and simulation results on controlling TCP rates by buffer management. We present a buffer management policy that provides loose rate guarantees to SACK TCP sources when the total buffer allocation is low. We study the performance of this buffer management scheme by simulation.

---

[1]This research was partially sponsored by the NASA Lewis Research Center under Grant Number NAS3-97198

# 1  Introduction: The Guaranteed Frame Rate Service

Guaranteed Frame Rate (GFR) has been recently proposed in the ATM Forum as an enhancement to the UBR service category. Guaranteed Frame Rate will provide a minimum rate guarantee to VCs at the frame level. The GFR service also allows for the fair usage of any extra network bandwidth. GFR requires minimum signaling and connection management functions, and depends on the network's ability to provide a minimum rate to each VC. GFR is likely to be used by applications that can neither specify the traffic parameters needed for a VBR VC, nor have cability for ABR (for rate based feedback control). Current internetworking applications fall into this category, and are not designed to run over QoS based networks. These applications could benefit from a minimum rate guarantee by the network, along with an opportunity to fairly use any additional bandwidth left over from higher priority connections. In the case of LANs connected by ATM backbones, network elements outside the ATM network could also benefit from GFR guarantees. For example, IP routers separated by an ATM network could use GFR VCs to exchange control messages. Figure 1 illustrates such a case where the ATM cloud connects several LANs and routers. ATM end systems may also establish GFR VCs for connections that can benefit from a minimum throughput guarantee.

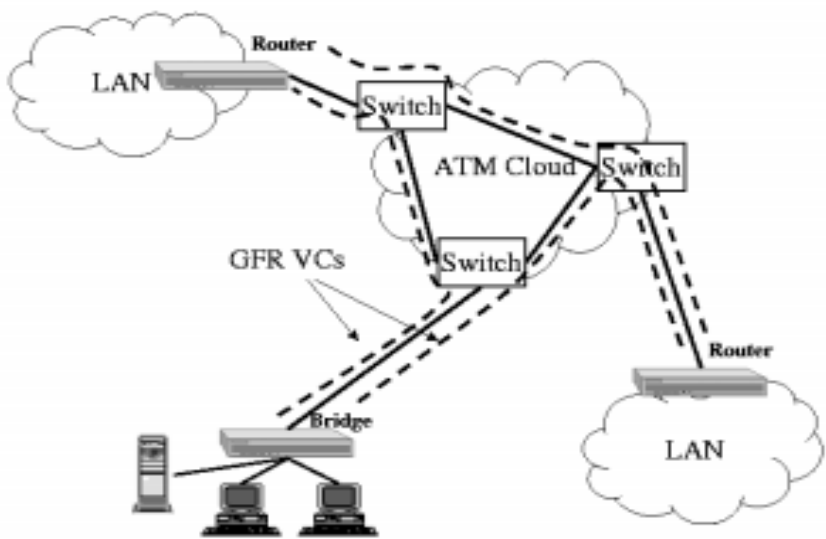

Figure 1: Use of GFR in ATM connected LANs

The original GFR proposals [11, 12] give the basic definition of the GFR service. GFR provides a minimum rate guarantee to the **frames** of a VC. The guarantee requires the specification of a maximum frame size (MFS) of the VC. If the user sends packets (or frames) smaller than the maximum frame size, at a rate less than the minimum cell rate (MCR), then all the packets are expected to be delivered by the network with minimum loss. If the user sends packets at a rate higher than the MCR, it should still receive at least the minimum rate. The minimum rate is guaranteed to the untagged frames of the connection. In addition, a connection sending in excess of the minimum rate should receive a fair share of any unused network capacity. The exact specification of the fair share has been left unspecified by the ATM Forum. Although the GFR specification is not yet finalized, the above discussion captures the essence of the service.

There are three basic design options that can be used by the *network* to provide the per-VC minimum rate guarantees for GFR – tagging, buffer management, and queueing:

1. **Tagging:** *Network based tagging* (or policing) can be used as a means of marking non-conforming packets before they enter the network. This form of tagging is usually performed when the connection enters the network. Figure 2 shows the role of network based tagging in providing a minimum rate service in a network. Network based tagging on a per-VC level requires some per-VC state information to be maintained by the network and increases the complexity of the network element. Tagging can isolate conforming and non-conforming traffic of each VC so that other rate enforcing mechanisms can use this information to schedule the conforming traffic in preference to non-conforming traffic. In a more general sense, policing can be used to discard non-conforming packets, thus allowing only conforming packets to enter the network.

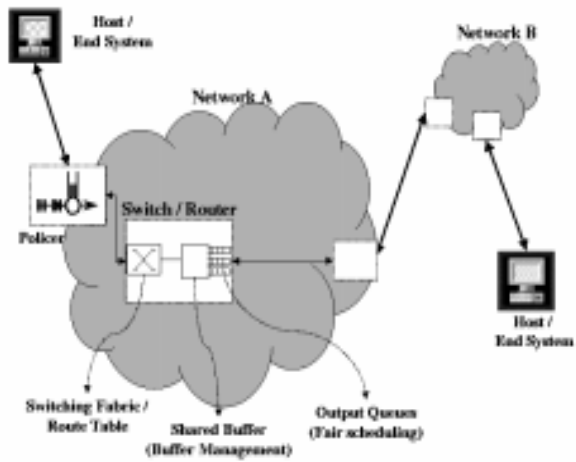

Figure 2: Network Architecture with tagging, buffer management and scheduling

2. **Buffer management:** Buffer management is typically performed by a network element (like a switch or a router) to control the number of packets entering its buffers. In a shared buffer environment, where multiple VCs share common buffer space, per-VC buffer management can control the buffer occupancies of individual VCs. Per-VC buffer management uses per-VC accounting to keep track of the buffer occupancies of each VC. Figure 2 shows the role of buffer management in the connection path. Examples of per-VC buffer management schemes are Selective Drop and Fair Buffer Allocation [9]. Per-VC accounting introduces overhead, but without per-VC accounting it is difficult to control the buffer occupancies of individual VCs (unless non-conforming packets are dropped at the entrance to the network by the policer). Note that per-VC buffer management uses a single FIFO queue for all the VCs. This is different from per-VC queuing and scheduling discussed below.

3. **Scheduling:** Figure 2 illustrates the position of scheduling in providing rate guarantees. While tagging and buffer management control the entry of packets into a network element, queuing strategies determine how packets are scheduled onto the next hop. FIFO queuing cannot isolate packets from various VCs at the egress of the queue. As a result, in a FIFO queue, packets are scheduled in the order in which they enter the buffer. Per-VC queuing, on the other hand, maintains a separate queue for each VC in the buffer. A scheduling mechanism can select between the queues at each scheduling time. However, scheduling adds the cost of per-VC queuing and the service discipline. For a simple service like GFR, this additional cost may be undesirable.

Several proposals have been made [3, 4, 8] to provide rate guarantees to TCP sources with FIFO queuing in the network. The bursty nature of TCP traffic makes it difficult to provide per-VC rate guarantees using FIFO queuing. Per-VC scheduling was recommended to provide rate guarantees to TCP connections. However, all these studies were performed at high target network utilization, i.e., most of the network buffers were allocated to the GFR VCs. We show that rate guarantees are achievable with a FIFO buffer for low buffer allocation.

All the previous studies have examined TCP traffic with a single TCP per VC. Per-VC buffer management for such cases reduces to per-TCP buffer management. However, routers that would use GFR VCs, would multiplex many TCP connections over a single VC. For VCs with several aggregated TCPs, per-VC control is unaware of each TCP in the VC. Moreover, aggregate TCP traffic characteristics and control requirements may be different from those of single TCP streams.

In this paper, we study two main issues:

- Providing minimum rate guarantees to TCP like adaptive traffic with FIFO buffer *for low rate allocations*.

- Buffer management of VCs with aggregate TCP flows.

Section 2 discusses the behavior of TCP traffic with controlled windows. This provides insight into controlling TCP rates by controlling TCP windows. Section 3 describes the effect of buffer occupancy and thresholds on TCP throughput. Section 4 presents a simple threshold-based buffer management policy to provide TCP throughputs in proportion to buffer thresholds for low rate allocations. This scheme assumes that each GFR VC may carry multiple TCP connections. We then present simulation results with TCP traffic over LANs interconnected by an ATM network. *In our simulation and analysis, we use SACK TCP [10] as the TCP model.*

## 2 TCP Behavior with Controlled Windows

TCP uses a window based mechanism for flow control. The amount of data sent by a TCP connection in one round trip is determined by the window size of the TCP connection. The window size is the minimum of the sender's congestion window (CWND) and the receiver's window (RCVWND). As a result, TCP rate can be controlled by controlling the window size of the TCP connection.

However, a window limit is not enforceable by the network to control the TCP rate. TCP sources respond to packet loss by reducing the source congestion window by one-half, and then increasing it by one segment size every round trip. As a result, the average TCP window can be controlled by packet discard at specific CWND values.

Figure 3 shows how the source TCP congestion window varies when a single segment is lost at a particular value of the congestion window. The figure is the CWND plot of the simulation of the configuration shown in Figure 4 with a single SACK TCP source (N=1). The figure shows four different values of the window at which a packet is lost. The round trip latency (RTT) for the connection is 30 ms. The window scale factor is used to allow the TCP window to increase beyond the 64K limit.

For window based flow control, the throughput (in Mbps) can be calculated from the average congestion window (in Bytes) and the round trip time (in seconds) as:

$$\text{Throughput (Mbps)} = \frac{8 \times 10^{-6} \times \text{Average CWND (bytes)}}{\text{Round Trip Time (secs)}} \quad (1)$$

The factor $8\times10^{-6}$ converts the throughput from bytes per sec to Megabits per sec. The average TCP CWND during the linear increase phase can be calculated as

$$\text{CWND}_{\text{avg}} = \frac{\Sigma_{i=1}^{T} \text{CWND}_{\max}/2 + \text{Max Segment Size} \times i}{T} \qquad (2)$$

where T is the number of round trip times for the congestion window to increase from $\text{CWND}_{\max}/2$ to $\text{CWND}_{\max}$. Note that this equation assumes that during the linear increase phase, the TCP window increases by one segment every round trip time. However, when the TCP delayed acknowledgment option is set, TCP might only send an ACK for every two segments. In this case, the window would increase by 1 segment every 2 RTTs.

From Figure 3, the average congestion windows in the linear phases of the four experiments are approximately 91232 bytes, 181952 bytes, 363392 bytes and over 600000 bytes. As a result, the average calculated throughputs from equation 1 are 24.32 Mbps, 48.5 Mbps, 96.9 Mbps, and 125.6 Mbps (126 Mbps is the maximum possible TCP throughput for a 155.52 Mbps link with 1024 byte TCP segments). The respective throughputs obtained from the simulations of the four cases are 23.64 Mbps, 47.53 Mbps, 93.77 Mbps and 25.5 Mbps. The throughput values calculated from the average congestion windows are close to those obtained by simulation. This shows that controlling the TCP window so as to maintain a desired average window size enables the network to control the average TCP throughput.

## 3  TCP Window Control using Buffer Management

In the previous section, an artificial simulation was presented where the network controlled the TCP rate by dropping a packet every time the TCP window reached a particular value. In practice, the ATM network knows neither the size of the TCP window, nor the round trip time of the connection. A switch can use per-VC accounting of the TCP packets in its buffer to estimate the bandwidth used by the connection.

In a FIFO buffer, the output rate of a connection is determined by the number of packets of the connection in the buffer. Let $\mu_i$ and $x_i$ be the output rate and the buffer occupancy respectively of $VC_i$. Let $\mu$ and $x$ be the total output rate and the buffer occupancy of the FIFO buffer respectively. Then, by the FIFO principle, in steady state,

$$\mu_i = \frac{x_i}{x}\mu$$

$$\text{or } \frac{x_i/x}{\mu_i/\mu} = 1$$

If the buffer occupancy of every active VC is maintained at a desired threshold, then the output rate of each VC can also be controlled. In other words, if a VC always has $x_i$ cells in the buffer with a total occupancy of $x$ cells, its average output rate will be at least $\mu x_i/x$.

Adaptive flows like TCP respond to segment loss by reducing their congestion window. A single packet loss is sufficient to reduce the TCP congestion window by one-half. Consider a drop policy that drops a single TCP packet from a connection every time the connection's buffer occupancy crosses a given threshold. The drop threshold for a connection determines the maximum size to which the congestion window is allowed to grow. Because of TCP's adaptive nature, the buffer occupancy reduces after about 1 RTT. The drop policy drops a single packet when the TCP's buffer occupancy crosses the threshold, and then allows the buffer occupancy to grow by accepting the remainder of the TCP window. On detecting a loss, TCP reduces its congestion window by 1 segment and remains idle for about one-half RTT, during which the buffer occupancy decreases below the threshold. Then the TCP window increases linearly (and so does the buffer occupancy), and a packet is again dropped when the buffer occupancy crosses the threshold. In this way, TCP windows can be controlled quite accurately to within one round trip time. As a result, the TCP's throughput can also be controlled by controlling the TCP's buffer occupancy.

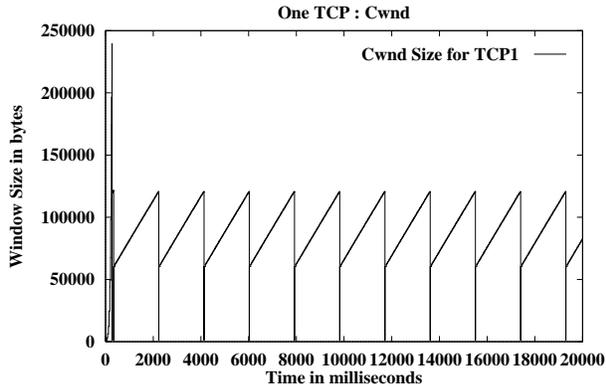

(a)

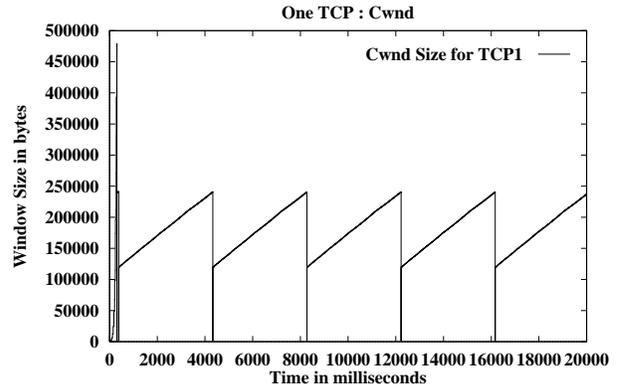

(b)

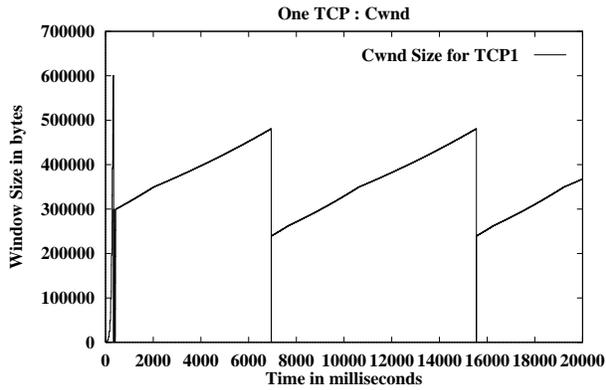

(c)

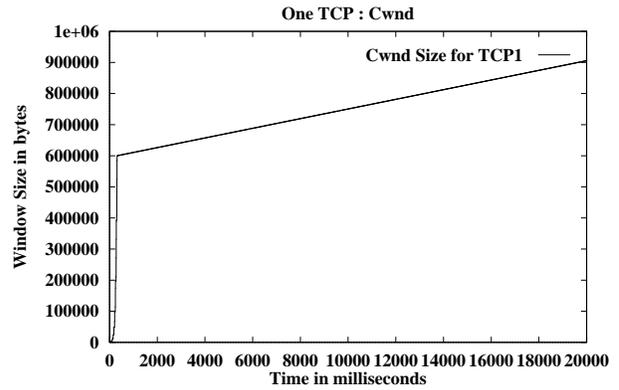

(d)

Figure 3: Single TCP Congestion Window Control. Drop thresholds (bytes) = 125000, 250000, 500000, None

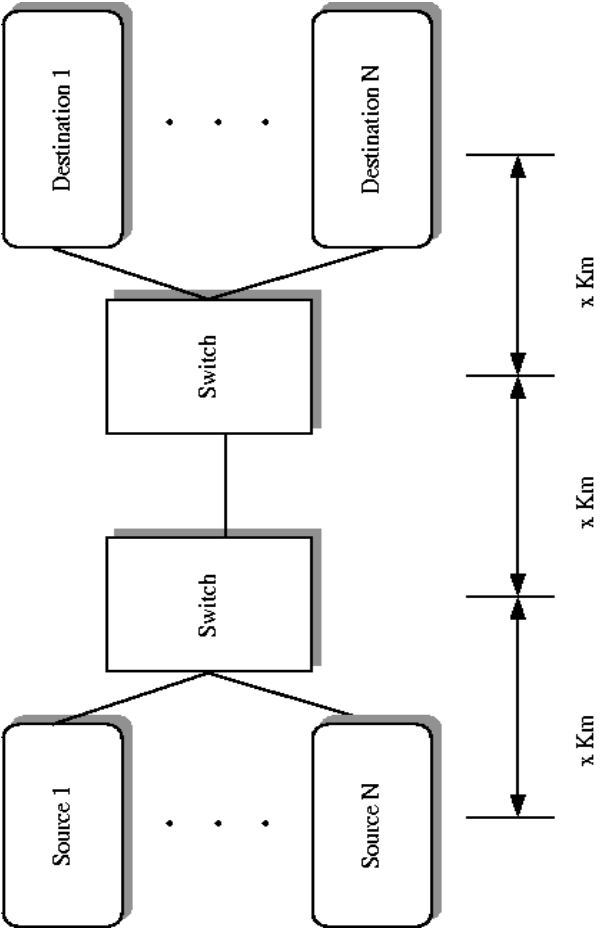

Figure 4: N source configuration

Table 1: Fifteen TCP buffer thresholds

| Experiment number | 1 | 2 | 3 | 4 |
|---|---|---|---|---|
| TCP number | Threshold per TCP (cells) ($r_i$) | | | |
| 1-3 | 305 | 458 | 611 | 764 |
| 4-6 | 611 | 917 | 1223 | 1528 |
| 7-9 | 917 | 1375 | 1834 | 2293 |
| 10-12 | 1223 | 1834 | 2446 | 3057 |
| 13-15 | 1528 | 2293 | 3057 | 3822 |
| Total threshold ($r$) | 13752 | 20631 | 27513 | 34392 |

Using this drop policy, we performed simulations of the TCP configuration in Figure 4 with fifteen TCP sources divided into 5 groups of 3 each. Each TCP source was a separate UBR VC. Five different buffer thresholds ($r_i$) were selected, and each of three TCP's in a group had the same buffer threshold. Table 1 lists the buffer thresholds for the VC's in the FIFO buffer of the switches. We performed experiments with 4 different sets of thresholds as shown by the threshold columns. The last row in the table shows the total buffer allocated ($r = \Sigma r_i$) to all the TCP connections for each simulation experiment. The total buffer size was large (48000 cells) so that there was enough space for the buffers to increase after the single packet drop. For a buffer size of 48000 cells, the total target buffer utilizations were 29%, 43%, 57%, 71% in the 4 columns of table 1 respectively. The selected buffer thresholds determine the MCR achieved by each connection. For each connection, the ratios of the thresholds to the total buffer allocation should be proportional to the ratios of the achieved per-VC throughputs to the total achieved throughput. In other words, if $\mu_i$, $\mu$, $r_i$ and $r$ represent the per-VC achieved throughputs, total throughput, per-VC buffer

Table 2: Fifteen TCP throughputs

| Experiment number | 1 | 2 | 3 | 4 | |
|---|---|---|---|---|---|
| TCP number | Achieved throughput per TCP (Mbps) ($\mu_i$) | | | | Expected Throughput |
| | | | | | ($\mu_i^e = \mu \times r_i/\Sigma r_i$) |
| 1-3 | 2.78 | 2.83 | 2.95 | 3.06 | 2.8 |
| 4-6 | 5.45 | 5.52 | 5.75 | 5.74 | 5.6 |
| 7-9 | 8.21 | 8.22 | 8.48 | 8.68 | 8.4 |
| 10-12 | 10.95 | 10.89 | 10.98 | 9.69 | 11.2 |
| 13-15 | 14.34 | 13.51 | 13.51 | 13.93 | 14.0 |
| Total throughput ($\mu$) | 125.21 | 122.97 | 125.04 | 123.35 | 126.0 |

Table 3: Fifteen TCP buffer:throughput ratio

| Experiment number | 1 | 2 | 3 | 4 |
|---|---|---|---|---|
| TCP number | Ratio ($\mu_i/\mu_i^e$) | | | |
| 1-3 | 1.00 | 1.03 | 1.02 | 1.08 |
| 4-6 | 0.98 | 1.01 | 1.03 | 1.04 |
| 7-9 | 0.98 | 1.00 | 1.00 | 1.02 |
| 10-12 | 0.98 | 0.99 | 0.98 | 0.88 |
| 13-15 | 1.02 | 0.98 | 0.97 | 1.01 |

allocations, and total buffer allocation respectively, then we should have

$$\mu_i/\mu = r_i/r$$

or the expected per-VC throughput is

$$\mu_i^e = \mu \times r_i/r$$

Table 2 shows the average throughput obtained per TCP in each group for each of the four simulations. The TCP throughputs were averaged over each group to reduce the effects of randomness. The last row of the table shows the total throughput obtained in each simulation. Based on the TCP segment size (1024 bytes) and the ATM overhead, it is clear that the TCPs were able to use almost the entire available link capacity (approximately 126 Mbps at the TCP layer).

The proportion of the buffer usable by each TCP ($r_i/r$) before the single packet drop should determine the proportion of the throughput achieved by the TCP. Table 3 shows the ratios ($\mu_i/\mu_i^e$) for each simulation. All ratios are close to 1. This indicates that the TCP throughputs are indeed proportional to the buffer allocations. The variations (not shown in the table) from the mean TCP throughputs increased as the total buffer thresholds increased (from left to right across the table). This is because the TCPs suffered a higher packet loss due to the reduced room to grow beyond the threshold. Thus, high buffer utilization produced more variation in achieved rate (last column of Table 3), whereas in low utilization cases, the resulting throughputs were in proportion to the buffer allocations.

Figure 5 shows the congestion windows of one TCP from each group for each of the four simulations. The graphs illustrate that the behaviors of the TCP congestion windows are very regular in these cases. The average throughput achieved by each TCP can be calculated from the graphs using equations 1 and 2. An intersting observation is that for each simulation, the slopes of the graphs during the linear increase are approximately the same for each TCP, i.e., for a given simulation, the rate of increase of CWND is the same for all TCPs regardless of their drop thresholds. We know that TCP windows increase by 1 segment every round trip time. Thus, we can conclude that for a given simulation, TCPs sharing the FIFO buffer experience similar queuing delays regardless of the individual per-connection thresholds at which their packets are dropped. This is because, if all TCP's buffer occupancies are close to their respective thresholds ($r_i$), then when a packet arrives at the buffer, it is queued behind cells from $\Sigma(r_i)$ packets, regardless of the connection to which it belongs. Consequently, each TCP experiences the same average queuing delay.

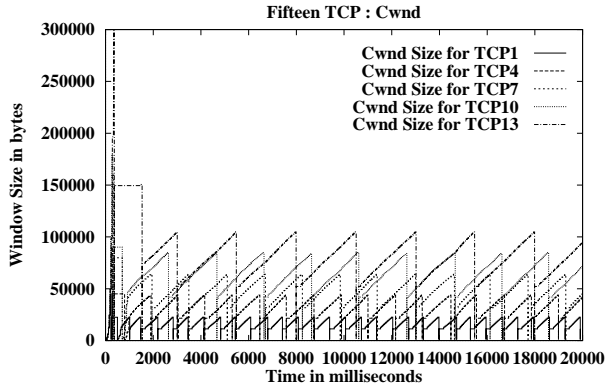

(a)

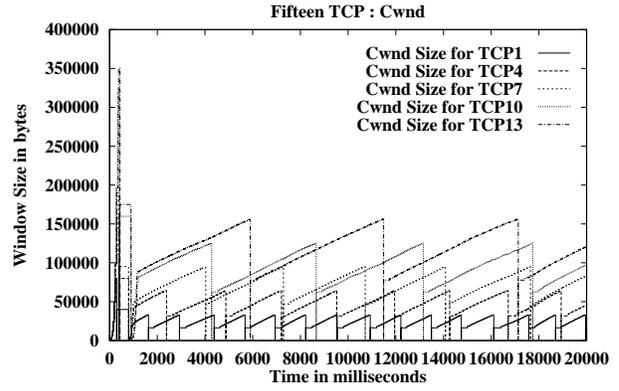

(b)

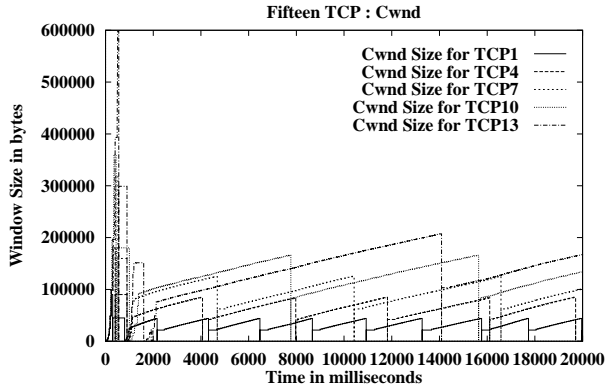

(c)

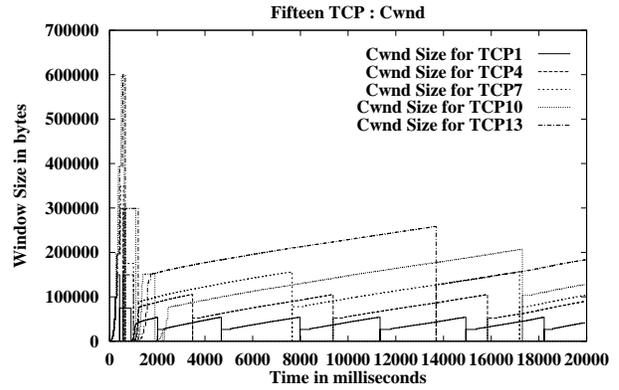

(d)

Figure 5: 15 TCP rate control by packet drop

However, as the total buffer threshold increases (from experiment (a) to (d)), the round trip time for each TCP increases because of the larger total queue size. The larger threshold also results in a larger congestion window at which a packet is dropped. A larger congestion window means that TCP can send more segments in one round trip time. But, the round trip time also increases proportionally to the increase in CWND (due to the increasing queuing delay of the 15 TCPs bottlenecked at the first switch). As a result, the average throughput achieved by a single TCP remains almost the same (see table 2) across the simulations. The formal proof of these conclusions will be presented in an extended version of this paper.

The following list summarizes the observations from the graphs:

1. TCP throughput can be controlled by controlling its congestion window, which in turn, can be controlled by setting buffer thresholds to drop packets. This statement clearly assumes that in cases where the offered load is low, and a queue is never built up, then the TCP is allowed to use as much capacity as it can.

2. With a FIFO buffer, the average throughput achieved by a connection is proportional to the fraction of the buffer occupancy of the connection's cells.

3. As long as the fraction of buffer occupancy of a TCP can be controlled, its relative throughput is independent of the total number of packets in the buffer, and depends primarily on the fraction of packets of that TCP in the buffer.

4. At a very high buffer utilization, packets may be dropped due to buffer unavailability. This results in larger variations in TCP throughputs. At very high thresholds, the queuing delay also increases significantly, and may cause the TCP sources to timeout.

5. At very low buffer thresholds (high loss rates), TCP sources become unstable and tend to timeout. Also, very low buffer occupancies result in low network utilization. Since TCP can maintain a flow of 1 CWND worth of packets each round trip time, a total buffer occupancy of 1 bandwidth-delay product should provide good utilization [13].

## 4 Buffer Management for GFR

In this section, we further develop the drop policy to design a buffer management scheme for the GFR service category. The goal of the scheme is to soft rate guarantees to SACK-TCP like adaptive traffic over ATM connections. The policy assumes that multiple TCP connections are multiplexed on a single VC. In this section we present the preliminary design and simulation results of the buffer management scheme. A parameter study and sensitivity analysis will be presented in a future study. Simulation results of heterogeneous TCP and non-TCP environments will be presented in a future study. *We assume a model in which TCPs may be merged into a single VC, in which case, the cells of different frames within a VC are not interleaved.* This allows the network to drop frames without having to identify the source that generated the frame.

Figure 6 illustrates a FIFO buffer for the GFR service category. The following attributes are defined:

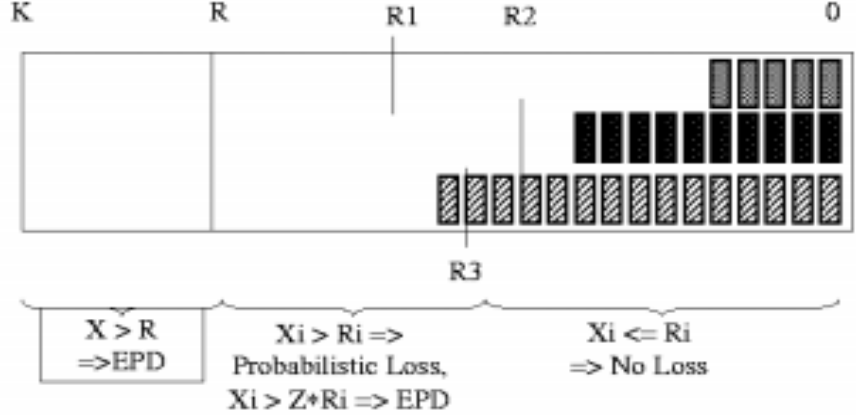

Figure 6: Drop behavior of Buffer Management scheme

- $K$: Buffer size in cells.
- $R$: Congestion threshold in cells ($0 \leq R \leq K$). EPD is performed when buffer occupancy is greater than $R$.
- $R_i$: Threshold for $VC_i$. (for example $R_i$ = function of ($\frac{MCR_i}{\text{Total UBR capacity}}$))
- $X$: Number of cells in the buffer.
- $X_i$: Number of cells of $VC_i$ in the buffer.
- $Z$: Scaling parameter for $R_i$ ($Z > 1$).
- $W_i$: Weight of $VC_i$ for probability calculation.
- $u$: Uniform(0,1) random number.

When the first cell of a frame arrives at the buffer, if the number of cells ($X_i$) of $VC_i$ in the buffer is less than its threshold ($R_i$) and if the total buffer occupancy $X$ is less than $R$, then the cell and frame is accepted into the buffer. If $X_i$ is greater than $R_i$, and if the total buffer occupancy ($X$) is greater than the buffer threshold ($R$), or if $X_i$ is greater than $Z \times R_i$, then the cell and frame are dropped (EPD). Thus $Z$ specifies a maximum per-VC buffer occupancy during congestion periods. Under low or mild load conditions, $R \times Z$ should be large enough to buffer a burst of cells without having to perform EPD. If the $X_i$ is greater than $R_i$, and $X$ is less than $R$, then the cell/frame are dropped in a probabilistic manner. The probability of frame drop depends on how much $X_i$ is above $R_i$, as well as the weight ($W_i$) of the connection. As $X_i$ increases beyond $R_i$, the probability of drop increases. Also, the drop probability should be higher for connections with a higher threshold. This is because, TCP flows with higher windows (due to higher thresholds) are more robust to packet loss than TCP flows with lower windows. Moreover, in the case of merged TCPs over a single VC, VCs with a high threshold are likely to carry more active TCP flows than those with a low threshold. As a result, a higher drop probability is more likely to hit more TCP sources and improve the fairness within a VC. $W_i$ is used to scale the drop probability according to desired level of control.

The frame is dropped with a probability

$$P\{drop\} = W_i \times \frac{X_i - R_i}{Z \times Ri - Ri}$$

In addition, if $X_i$ is greater than $R_i$, then all tagged frames may also be dropped. Tagging support is not yet tested for this drop policy.

The resulting algorithm works as follows. When the first cell of a frame arrives:

```
IF ((Xi < Ri AND X < R)) THEN

        ACCEPT CELL AND REMAINING CELLS IN FRAME

ELSE IF  ((X < R) AND (Xi < Z*Ri)  AND
          (Cell NOT Tagged)  AND
          (u > Wi*(Xi - Ri)/(Ri(Z-1))))

     THEN ACCEPT CELL AND REMAINING CELLS IN FRAME

     ELSE DROP CELL AND REMAINING CELLS IN FRAME

ENDIF
```

If the bufer occupancy exceeds the total buffer size, then, the cell must be dropped. In this case partial packet discard is performed.

Figure 7 illustrates the 15 TCP configuration in which groups of three TCPs are merged into 1 single VC. Each local switch (edge device separating the LAN from the backbone ATM network) merges the 3 TCPs into a single GFR VC over the backbone link. The backbone link has 5 VCs going through it, each with 3 TCPs. The local switches ensure that the cells of frames within a single VC are not interleaved. The backbone switches implement the buffer allocation policy described above. The local switches are not congested in this configuration.

We simulated the 15 merged TCP configuration with 3 different buffer threshold sets. The parameter $Z$ was set to 1.5, therefore, EPD was performed for each VC when its buffer occupancy was $1.5 \times R$. Table 4 shows the thresholds used for each VC at the first bottleneck switch.

Table 5 shows the ratio $(\mu_i/\mu)/(r_i/\Sigma r_i)$ for each VC for the configuration in Figure 7 and the corresponding thresholds. In all cases, the achieved link utilization was almost 100%. The table shows that TCP throughputs obtained were in proportion to the buffers allocated (since most of the ratios in table 5 are close to 1). The highest variation (not shown in the table) was seen in the last column because of the high threshold values.

In our simulations, the maximum observed queue sizes in cells in the first backbone switch (the main bottleneck) were 3185, 5980 and 11992 respectively. The total allocated buffer thresholds were 2230, 4584 and 9171 cells for the experiments. At higher buffer allocations, the drop policy cannot provide tight bounds on throughput.

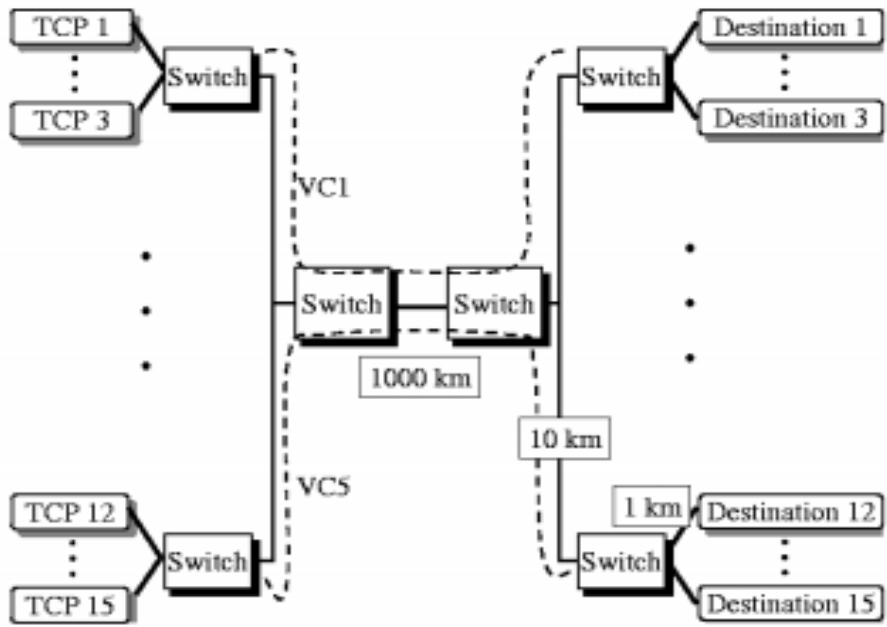

Figure 7: N source VC merge configuration

Table 4: Differential FBA thresholds

| VC number | Threshold (cells) | | |
|---|---|---|---|
| 1 | 152 | 305 | 611 |
| 2 | 305 | 611 | 1223 |
| 3 | 458 | 917 | 1834 |
| 4 | 611 | 1223 | 2446 |
| 5 | 764 | 1528 | 3057 |
| Total | 2290 | 4584 | 9171 |

## 5  Summary and Future Work

In this paper, we have used FIFO buffers to control SACK TCP rates by buffer management. An optimal set of thresholds should be selected that is high enough to provide sufficient network utilization, and is low enough to allow stable operation. The achieved TCP throughputs are in proportion to the fraction of the average buffer occupied by the VC.

More work remains to be done to further modify the buffer management scheme to work with a variety of configurations. In particular, we have only studied the performance of this scheme with SACK TCP. Its performance with heterogeneous TCPs is a topic of further study. We have not studied the effect of non adaptive traffic (like UDP) on the drop policy. It appears that for non adaptive traffic, the thresholds must be set lower than those for adaptive traffic (for the same MCR), and the dropping should be more strict when the buffer occupancy crosses the threshold. In this paper we have not studied the effect of network based tagging in the context of GFR. In the strict sense, GFR only provides a low CLR guarantee to the CLP=0 cell stream i.e., the cells that were not tagged by the source and passed the GCRA conformance test. However, when source (this could be a non-ATM network element like a router) based tagging is not performed, it is not clear if the CLP0 stream has any significance over the CLP1 stream. Moreover,

Table 5: Differential FBA simulation results

| VC number | Ratio $(\mu_i/\mu)/(r_i/r)$ | | |
|---|---|---|---|
| 1 | 1.04 | 1.01 | 1.16 |
| 2 | 1.05 | 1.02 | 1.06 |
| 3 | 0.97 | 0.99 | 1.05 |
| 4 | 0.93 | 1.00 | 1.13 |
| 5 | 1.03 | 0.99 | 0.80 |

network tagging is an option that must be signaled during connection establishment.

---

[2] All our papers and ATM Forum contributions are available from http://www.cis.ohio-state.edu/~jain